\documentclass[runningheads]{llncs}
\usepackage{graphicx}
\usepackage{amsmath}

\DeclareMathOperator*{\argmin}{arg\,min}
\begin{document}
\title{Temporal Feature Fusion with Sampling Pattern Optimization for Multi-echo Gradient Echo Acquisition and Image Reconstruction}
%
%
\author{Jinwei Zhang\inst{1,3} \and
Hang Zhang\inst{2,3} \and
Chao Li\inst{3,4} \and
Pascal Spincemaille\inst{3} \and
Mert Sabuncu\inst{1,2,3} \and
Thanh D. Nguyen\inst{3} \and
Yi Wang\inst{1,2,3}}

\authorrunning{J. Zhang et al.}
%
\institute{Department of Biomedical Engineering, Cornell University, Ithaca, NY, USA \and
Department of Electrical and Computer Engineering, Cornell University, Ithaca, NY, USA \and
Department of Radiology, Weill Medical College of Cornell University, New York, NY, USA \and
Department of Applied Physics, Cornell University, Ithaca, NY, USA}
%
%
%
\maketitle              
\begin{abstract}
Quantitative imaging in MRI usually involves acquisition and reconstruction of a series of images at multi-echo time points, which possibly requires more scan time and specific reconstruction technique compared to conventional qualitative imaging. In this work, we focus on optimizing the acquisition and reconstruction process of multi-echo gradient echo pulse sequence for quantitative susceptibility mapping as one important quantitative imaging method in MRI. A multi-echo sampling pattern optimization block extended from LOUPE-ST is proposed to optimize the k-space sampling patterns along echoes. Besides, a recurrent temporal feature fusion block is proposed and inserted into a backbone deep ADMM network to capture the signal evolution along echo time during reconstruction. Experiments show that both blocks help improve multi-echo image reconstruction performance.

\end{abstract}
\section{Introduction}
Quantitative imaging (QI) is an emerging technology in magnetic resonance imaging (MRI) to quantify tissues' magnetic properties. Typical QI methods involve tissues' parameter mapping such as T1 and T2 relaxation time quantification \cite{deichmann2005fast,deoni2005high}, water/fat separation with R2$^*$ estimation \cite{yu2008multiecho}, quantitative susceptibility mapping (QSM) \cite{wang2015quantitative} and etc. These parameters provide new biomarkers for the clinical assessment of diverse diseases. In QI, a multi-echo pulse sequence is used to acquire MRI signals at different echo times. After image reconstruction for all echo times, a temporal evolution model with respect to the parameters of interest is used to compute the parameters via a model-based nonlinear least square fitting. For example, QSM estimates the tissue induced local magnetic field by a nonlinear fitting of complex multi-echo signals \cite{liu2013nonlinear} and solves tissue susceptibility using regularized dipole inversion \cite{liu2012morphology}.

Despite its merits for tissue parameter quantification, QI has a major drawback of extended scan time compared to conventional qualitative imaging, as the echo time needs to be long enough to cover the temporal evolution for all tissue types in order to obtain accurate parameter mapping. Classical acceleration techniques in MRI can be incorporated into QI. Parallel imaging (PI) \cite{griswold2002generalized,pruessmann1999sense}, compressed sensing (CS) \cite{lustig2007sparse} and their combinations (PI-CS) \cite{murphy2012fast,otazo2010combination} are widely used techniques for acquiring and reconstructing under-sampled k-space data to shorten scan time, which is referred to as under-sampled k-p space acquisition and reconstruction in QI \cite{zhang2015accelerating,zhao2015accelerated}. To acquire k-p space data, a variable density under-sampling pattern at each echo is applied, where the design of the ‘optimal’ under-sampling patterns across echoes remains an open problem in QI. In fact, various methods have been proposed for single-echo sampling pattern optimization, which may inspire multi-echo optimal sampling pattern design for QI. Representative single-echo sampling pattern optimization methods include machine learning based optimization LOUPE \cite{bahadir2020deep} and its extension LOUPE-ST \cite{zhang2020extending}, experimental design with the constrained Cramer-Rao bound OEDIPUS \cite{haldar2019oedipus} and greedy pattern selection \cite{gozcu2018learning}.

After k-p space acquisition, the next step is to reconstruct multi-echo images from under-sampled data. Based on the observation that correlation exists among multi-echo image structures, simultaneous reconstruction of multi-echo images with joint sparsity regularization was introduced for QI \cite{zhao2015accelerated}. Besides, globally low rank Casorati matrix was imposed assuming the profiles of signal decay curves at all spatial locations were highly redundant \cite{peng2016accelerated}. Accordingly, a locally low rank regularization was applied by restricting the Casorati matrix to different local image regions \cite{zhang2015accelerating}. To reconstruct multi-echo images using the above regularizations, these methods minimized the corresponding objective functions via an iterative scheme. Recently, with the advance of convolutional neural network, unrolling the whole iterative optimization process with a learned regularizer has become popular in under-sampled k-space reconstruction. Pioneering methods include MoDL \cite{aggarwal2018modl} and VarNet \cite{hammernik2018learning} for single-echo image reconstruction and cascaded \cite{schlemper2017deep} and recurrent network \cite{qin2018convolutional} for dynamic image sequence reconstruction, which also inspire the design of unrolled reconstruction network for multi-echo QI.

In this paper, we focus on quantitative susceptibility mapping (QSM) \cite{wang2015quantitative} acquired with Multi-Echo GRadient Echo (MEGRE) pulse sequence. In QSM, MEGRE signals are generated based on the following signal evolution model:
\begin{equation}
 s_j = m_0 e^{-R_2^*t_j} e^{i(\phi_0+f\cdot t_j)},
 \label{signal_model}
\end{equation}
where $s_j$ is the signal at the $j$-th echo time $t_j$, $m_0$ is the proton density (water) in the tissue, $R_2^*$ is the effective transverse magnetization relaxation factor, $\phi_0$ is the initial phase at radio-frequency (RF) pulse excitation, and $f$ is the total magnetic field data generated by the tissue and air susceptibility. After acquiring and reconstructing $\{s_j\}$ of all echoes, a nonlinear field estimation with Levenberg-Marquardt algorithm \cite{liu2013nonlinear} is used to estimate $f$ from Eq. \ref{signal_model}, then a morphology enabled dipole inversion (MEDI) method \cite{liu2012morphology} is deployed to compute the corresponding susceptibility map from $f$. In this work, we attempt to combining the optimization of acquisition and reconstruction process of MEGRE signals into one learning based approach. We extend LOUPE-ST \cite{zhang2020extending} to the multi-echo scenario for temporal sampling pattern optimization and propose a novel temporal feature fusion block into a deep ADMM based unrolled reconstruction network to capture MEGRE signal evolution during reconstruction.


\section{Method}
The mulit-coil multi-echo k-space under-sampling process is as follows:
\begin{equation}
    b_{jk} = U_jFE_ks_j + n_{jk},
    \label{kspace}
\end{equation}
where $b_{jk}$ is the measured under-sampled k-space data of the $k$-th receiver coil at the $j$-th echo time with $N_C$ receiver coils and $N_T$ echo times in total, $U_j$ is the k-space under-sampling pattern at $j$-th echo time, $F$ is the Fourier transform, $E_k$ is the sensitivity map of $k$-th coil, $s_j$ is the complex image of the $j$-th coil to be reconstructed, and $n_{jk}$ is the voxel-by-voxel i.i.d. Gaussian noise when measuring $b_{jk}$. Notice that $U_j$ may vary across echoes, which provides the flexibility of sampling pattern design for different echo time.

Having acquired $\{ b_{jk}\}$ with fixed $\{ U_j \}$, we aim at reconstructing $\{s_j\}$ of all echoes simultaneously with a cross-echo regularization $R(\{s_j\})$. Based on Eq. \ref{kspace}, the objective function to minimize is:
\begin{equation}
    E(\{s_j\}; \{ U_j\}) = R(\{s_j\}) + \sum_{j,k}\| U_jFE_ks_j - b_{jk} \|_2^2.
    \label{objective}
\end{equation}
In the following sections, we will minimize Eq. \ref{objective} iteratively with a learned regularizer $R(\{s_j\})$ which captures the dynamic evolution along echoes. We will also design $\{U_j\}$ to boost the reconstruction performance under a fixed Cartesian under-sampling ratio.

\begin{figure}[t!]
  {\includegraphics[width=\textwidth]{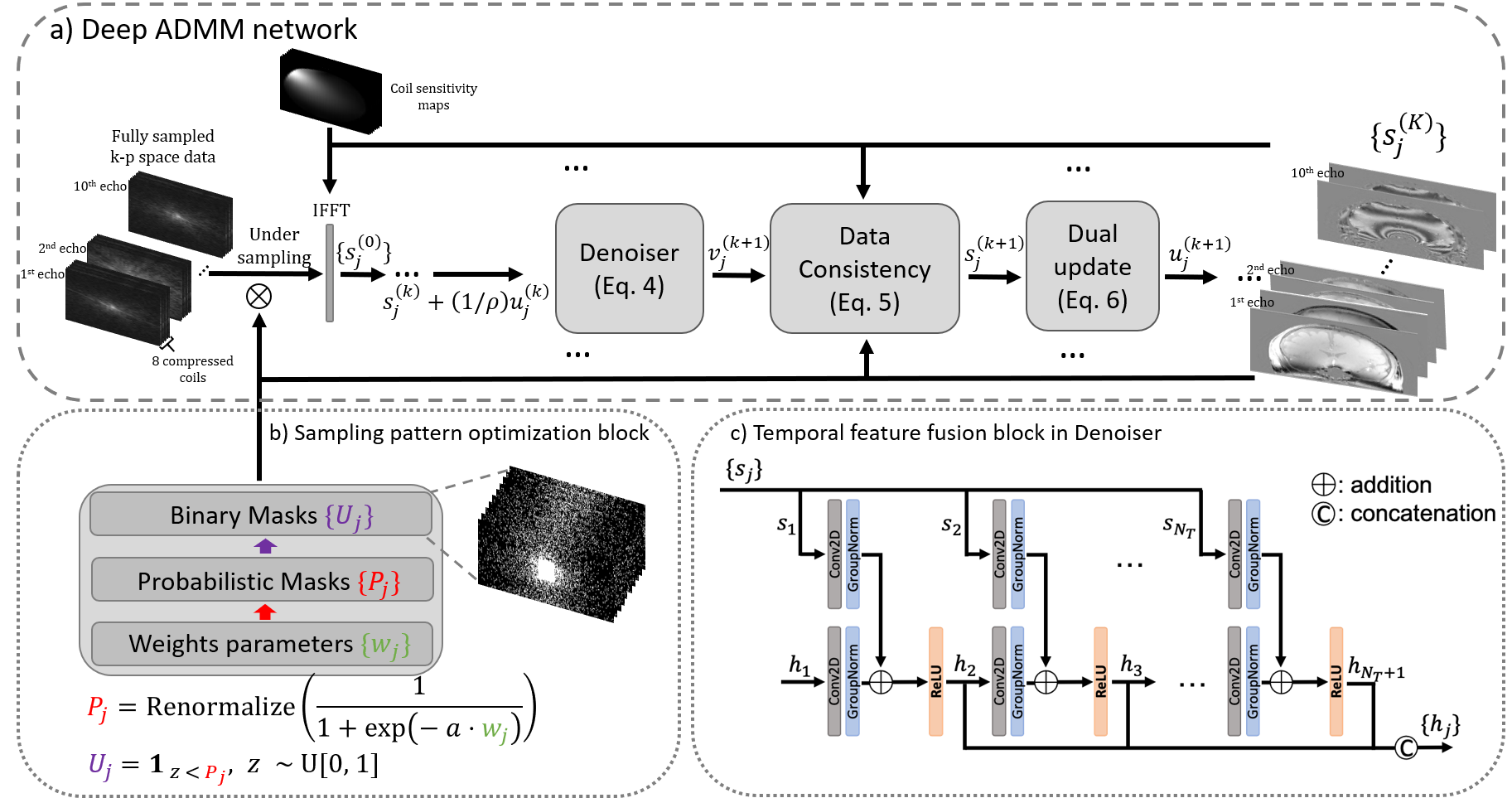}}
  {\caption{Network architecture. Deep ADMM (a) was used as backbone for under-sampled k-p space reconstruction. A sampling pattern optimization block (b) extended from LOUPE-ST was used to learn optimal multi-echo patterns. A temporal feature fusion block (c) was inserted into deep ADMM to capture signal evolution along echoes.}}
  \label{fig2}
\end{figure}

\subsection{Deep ADMM as backbone}
Assuming $U_j$'s are pre-defined, we use alternating direction method of multiplier (ADMM) \cite{boyd2011distributed} to minimize Eq. \ref{objective}. Introducing auxiliary variable $v_j = s_j$ for all $j$'s, ADMM splits the original problem into a sequence of subproblems:
\begin{align}
    & \{v_j^{(k+1)}\} = \argmin_{\{v_j\}} R(\{v_j\}) + \frac{\rho}{2} \sum_j \| v_j - \Tilde{v}_j^{(k)}\|_2^2, \\
    & \{s_j^{(k+1)}\} = \argmin_{\{s_j\}} \sum_{j,k}\| U_jFE_ks_j - b_{jk} \|_2^2 + \frac{\rho}{2} \sum_j \| s_j - \Tilde{s}_j^{(k)}\|_2^2, \\
    & \{u_j^{(k+1)}\} = \{u_j^{(k)}\} + \rho (\{s_j^{(k+1)}\} - \{v_j^{(k+1)}\}),
\end{align}
where $\Tilde{v}_j^{(k)} = s_j^{(k)} + \frac{1}{\rho}u_j^{(k)}$, $\Tilde{s}_j^{(k)} = v_j^{(k+1)} - \frac{1}{\rho}u_j^{(k)}$, $u_j^{(k)}$ is the dual variable and $\rho$ is the penalty parameter in ADMM. To choose a regularizer $R(\{s_j\})$ in Eq. 4, a plug-and-play ADMM \cite{chan2016plug} strategy was proposed to replace the "denoising" step Eq. 4 by an off-the-shelf image denoising algorithm $\{v_j^{(k+1)}\} = \mathcal{D}(\{\Tilde{v}_j^{(k)}\})$ for flexibility and efficiency. Inspired by a recent work MoDL \cite{aggarwal2018modl} where an MR image reconstruction network was built by unrolling the quasi-Newton iterative scheme and learning the regularizer from fully-sampled data, we propose to unroll the iterative schemes in Eq. 4-6 to a data graph and design $\mathcal{D}(\{\Tilde{v}_j^{(k)}\})$ as a convolutional neural network (CNN) to be trained from data. The whole reconstruction network architecture is shown in Fig. 1(a), which we call deep ADMM. Multi-echo real and imaginary parts of MR images are concatenated into the channel dimension, yielding $2N_T$ channels in $\mathcal{D(\cdot)}$. In the backbone deep ADMM, vanilla $\mathcal{D(\cdot)}$ consists of five convolutional layers with zero padding to preserve image spatial dimension. Data consistency subproblem Eq. 5 is solved by conjugate gradient (CG) descent algorithm. Both weights in $\mathcal{D(\cdot)}$ and 
penalty parameter $\rho$ are learnable parameters in deep ADMM.

\subsection{Temporal feature fusion block}
The vanilla regularizer $\mathcal{D(\cdot)}$ of deep ADMM proposed in section 2.1 aggregates cross-echo information through channel dimension, which may not capture the dynamic signal evolution along time in Eq. \ref{signal_model}. In order to take into account the signal evolution based regularization, we propose a temporal feature fusion (TFF) block as shown in Fig. 1(c). In TFF, a recurrent module is repeated $N_T$ times with shared weights, in which at the $j$-th repetition single echo image $s_j$ and hidden state feature $h_j$ are fed into the module to generate next hidden state feature $h_{j+1}$. Since $h_j$ is meant to carry information from preceding echoes, the concatenated multi-echo hidden states $\{h_j\}$ is able to capture the echo evolution and fuse features across all echoes, which will then be fed into a denoising block to generate denoised multi-echo images. The signal evolution model along echo time is implicitly incorporated during the recurrent feed forward process due to the parameter sharing mechanism which effectively leverages the relationship between input of one echo and its temporal neighbors.

\subsection{Sampling pattern optimization block}
Another goal is to design optimal k-p space sampling pattern $\{U_j\}$ for better image reconstruction performance. In this work, we focus on 2D variable density sampling pattern with a fixed under-sampling ratio. We accomplish this by extending single-echo LOUPE-ST \cite{zhang2020extending} to the multi-echo scenario as our sampling pattern optiomization (SPO) block. The multi-echo LOUPE-ST block is shown in Fig. 1(b). Learnable weights $\{w_j\}$ with echo index $j$ are used to generate multi-echo probabilistic patterns $\{P_j\}$ through sigmoid transformation and sampling ratio renormalization. Binary multi-echo under-sampling patterns $\{ U_j \}$ are generated via stochastic sampling from $\{P_j\}$, i.e., $U_j = \textbf{1}_{z < P_j}$ for each echo with $\textbf{1}_x$ the indicator function on the truth value of $x$ and $z$ uniformly distributed between [0, 1]. Then $\{ U_j \}$ are used to retrospectively acquire $\{ b_{jk}\}$ from fully-sampled multi-echo k-p space dataset.
LOUPE-ST applied a straight-through estimator \cite{bengio2013estimating} for back-propagation to the stochastic sampling layer where binary pattern was generated, which was reported to perform better than the vanilla LOUPE \cite{bahadir2020deep}. We refer readers to \cite{zhang2020extending} for details.

\section{Experiments}
\begin{table}[t!]
\centering
\caption{Ablation study. Reconstruction performances were progressively improved as more blocks were added to deep ADMM (* denotes statistical significance; $p < 0.05$).}
\begin{tabular}{c|c|c} \hline
  & PSNR ($\uparrow$) & SSIM ($\uparrow$) \\ 
\hline
Deep ADMM  & 40.95 $\pm$ \ 3.72* & 0.9820 $\pm$ \ 0.0257*\\
Deep ADMM + single SPO  & 41.77 $\pm$ \ 3.24* & 0.9839 $\pm$ \ 0.0091*\\
Deep ADMM + TFF & 42.24 $\pm$ \ 3.28* & 0.9864  $\pm$ \ 0.0074*\\
Deep ADMM + TFF + single SPO & 42.77 $\pm$ \ 3.38* & 0.9867 $\pm$ \ 0.0076*\\
Deep ADMM + TFF + multi SPO  & $\textbf{43.75} \ \pm$ \ 3.02 & $\textbf{0.9894} \ \pm$ \ 0.0058\\
\hline
\end{tabular}
\label{tab1}
\end{table}

\subsection{Data acquisition and preprocessing}
Cartesian fully sampled k-space data of multi-echo images were acquired in 7 subjects using a 3D MEGRE sequence on a 3T GE scanner with a 32-channel head coil. Imaging parameters were: $256\times206\times80$ imaging matrix size with the corresponding $\text{readout} \times \text{phase} \times \text{phase}$ encoding directions, $1\times1\times2 \ \text{mm}^3$ resolution, 10 echoes with 1.972 ms as the first TE and 3.384 ms echo spacing. 32-coil k-space data of each echo were compressed into 8 virtual coils using a geometric singular value decomposition based coil compression algorithm \cite{zhang2013coil}. After compression, coil sensitivity maps of the first echo were estimated with a reconstruction null space eigenvector decomposition based algorithm ESPIRiT \cite{uecker2014espirit} using a centric $24\times24\times24$ auto-calibration k-space region for each compressed coil. The estimated coil sensitivity maps were also used for the remaining echoes during reconstruction. From the fully sampled k-space data, coil combined multi-echo images were computed to provide the ground truth labels for both network training and performance comparison. Central 200 slices along readout direction containing brain tissues of each fully sampled subject were extracted and 3/1/3 subjects (600/200/600 slices) were used as training, validation and test datasets.

\subsection{Implementation details and ablation study}
\begin{table}[t!]
\centering
\caption{Performance comparison. For each sampling pattern, Deep ADMM + TFF performed better than LLR and multi-echo MoDL (* denotes statistical significance; $p < 0.05$). For each reconstruction method, improvements were observed from manually designed to multi-echo SPO patterns.}
\begin{tabular}{c|c|c|c|c} \hline
& \multicolumn{2}{c}{Variable density} & \multicolumn{2}{c}{Single SPO} \\
\cline{2-5}
& PSNR ($\uparrow$) & SSIM ($\uparrow$) & PSNR ($\uparrow$) & SSIM ($\uparrow$) \\ 
\hline
LLR  & 36.68 $\pm$ \ 3.70* & 0.9604 $\pm$ \ 0.0230* & 36.72 $\pm$ \ 3.77* & 0.9582 $\pm$ \ 0.0248* \\
Multi-echo MoDL  & 40.45 $\pm$ \ 3.45* & 0.9809 $\pm$ \ 0.0101* & 41.01 $\pm$ \ 3.30* & 0.9817 $\pm$ \ 0.0099* \\
Deep ADMM + TFF & $\textbf{42.24} \ \pm$ \ 3.28 & $\textbf{0.9864} \ \pm$ \ 0.0074 & $\textbf{42.77} \ \pm$ \ 3.38 &  $\textbf{0.9867} \ \pm$ \ 0.0076 \\
\hline
\end{tabular}
\begin{tabular}{c|c|c} \hline
& \multicolumn{2}{c}{Multi SPO} \\
\cline{2-3}
& PSNR ($\uparrow$) & SSIM ($\uparrow$) \\ 
\hline
LLR  & 37.93 $\pm$ \ 3.65* & 0.9699 $\pm$ \ 0.0179* \\
Multi-echo MoDL  & 42.28 $\pm$ \ 3.02* & 0.9859 $\pm$ \ 0.0078* \\
Deep ADMM + TFF & $\textbf{43.75} \ \pm$ \ 3.02 & $\textbf{0.9894}  \ \pm$ \ 0.0058 \\
\hline
\end{tabular}
\label{tab2}
\end{table}

\begin{figure}[t!]
  {\includegraphics[width=\textwidth]{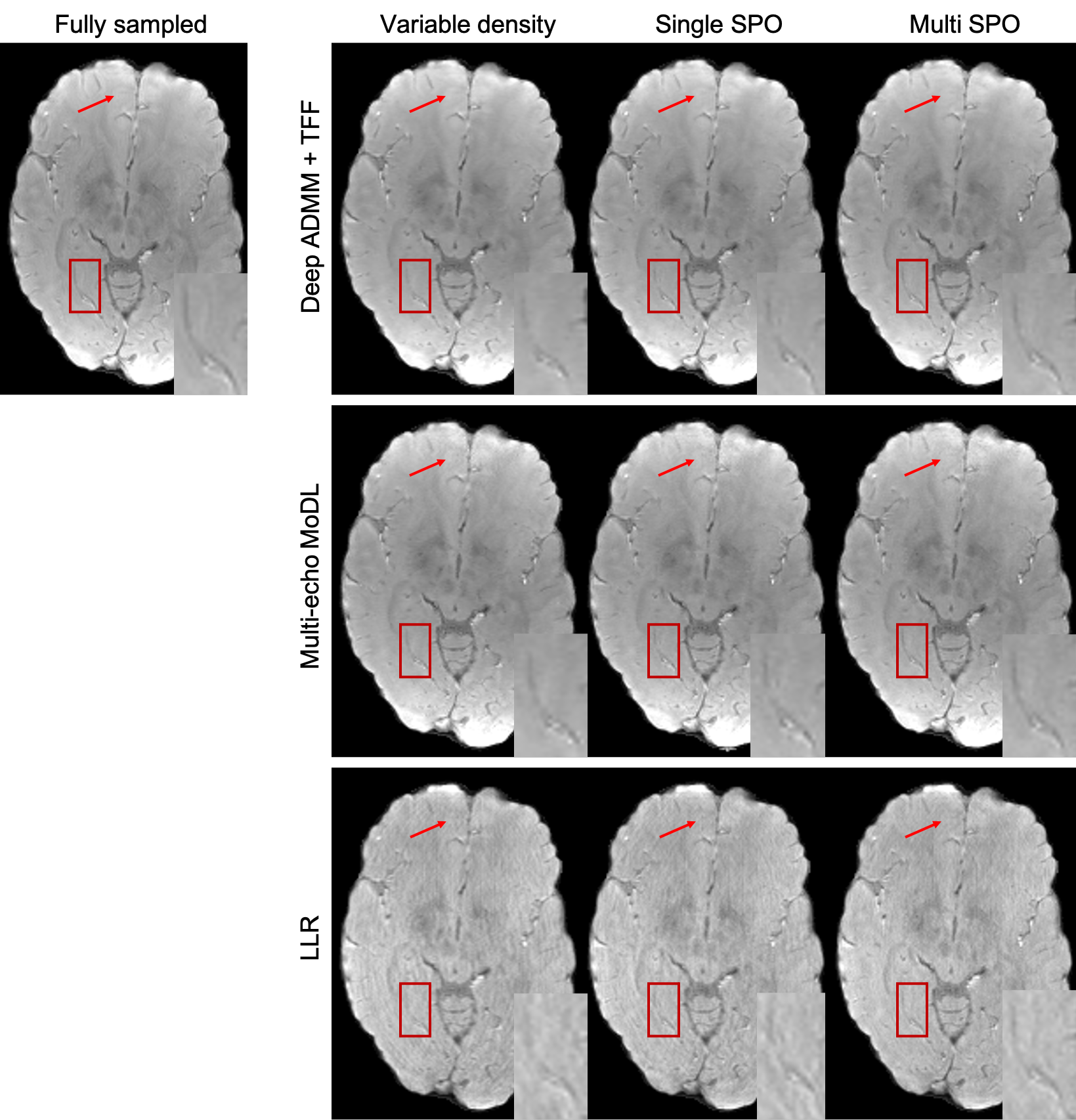}}
  {\caption{Performance comparison of one echo-combined test slice. Heavy artifacts were present almost everywhere in LLR, while mild artifacts (red arrows) were shown in multi-echo MoDL. In contrast, deep ADMM with TFF produced clean images without visible artifacts. For each method, zoom-in veil structures (from red boxes) were progressively improved from manually designed to multi-echo SPO patterns.}}
  \label{fig3}
\end{figure}

$23\%$ retrospective under-sampling on the coil-compressed fully sampled k-space data was applied in the experiment. The input 20-channel (real and imaginary parts with 10 echoes) zero-filled images were fed into deep ADMM with $K=10$ unrolled iterations to generate reconstructed images (Fig. 1a). During training, both weights in deep ADMM network and SPO block were updated simultaneously by minimizing a channel-wise structural similarity index measure (SSIM) \cite{wang2004image} loss: $\sum_{k=1}^K \sum_{j=1}^{N_T}\text{SSIM}(s_j^{(k)}, s^*_j)$, 
where $s^*_j$ denotes label and $s_j^{(k)}$ denotes $k$-th deep ADMM iteration at $j$-th echo. After training, a specific set of binary patterns $\{ U_j \}$ were generated from $\{ P_j \}$ and used to further train deep ADMM alone. We implemented in PyTorch using the Adam optimizer \cite{kingma2014adam} (batch size 1, number of epochs 100 and initial learning rate $10^{-3}$) on a RTX 2080Ti GPU. During test, the retrospectively under-sampled k-space data were fed into deep ADMM for inference. PSNR and SSIM on the echo-combined image $\sqrt{\sum_{j=1}^{N_T}|s_j|^2}$ were used to measure reconstruction quality.

An ablation study regarding TFF and SFO blocks were investigated and reconstruction performances on test dataset are shown in Table 1. A variable density sampling pattern was manually designed based on a multi-level sampling scheme \cite{roman2014asymptotic} and used to train baseline deep ADMM without TFF and SFO (first row in Table 1). TFF, single-echo SPO and multi-echo SPO were progressively added to the baseline deep ADMM (2-5 rows in Table 1) to check the effectiveness of each block. For Deep ADMM without TFF, the number of convolutional kernels in the denoiser block were extended to match the memory consumption to deep ADMM with TFF. In Table 1, reconstruction performance was progressively improved as more blocks were added to the baseline deep ADMM, where Deep ADMM with both TFF and multi-echo SPO blocks achieved the best performance. All the sampling patterns designed/learned in the ablation study are shown in the Appendix.

\subsection{Performance comparison}
The proposed deep ADMM + TFF network was compared with locally low rank (LLR) \cite{zhang2015accelerating} and multi-echo MoDL \cite{aggarwal2018modl}, where the vanilla MoDL was modified to reconstruct multi-echo image simultaneously. Manually designed variable density, learned single-echo SPO and multi-echo SPO patterns in the ablation study were used for performance comparison. Reconstructed echo-combined images of one test slice using different sampling patterns and reconstruction methods are shown in Fig. 2. Heavy artifacts were present almost everywhere in LLR, while mild artifacts (red arrows) were shown in multi-echo MoDL. In contrast, deep ADMM with TFF produced clean images without visible artifacts. For each method, zoom-in veil structures (from red boxes) were progressively improved from manually designed pattern to multi-echo SPO patterns. Quantitative metrics are shown in Table 2. Deep ADMM with TFF performed the best for each type of sampling pattern. Besides, for each reconstruction method, improvements were observed from manually designed, single SPO to multi-echo SPO patterns. QSMs were estimated from deep ADMM + TFF reconstructed multi-echo images and shown in Fig. 3 with the same slice to Fig. 2. The same veil structures which appeared bright in QSMs were also progressively improved from manually designed pattern to multi-echo SPO patterns.

\begin{figure}[t!]
  {\includegraphics[width=\textwidth]{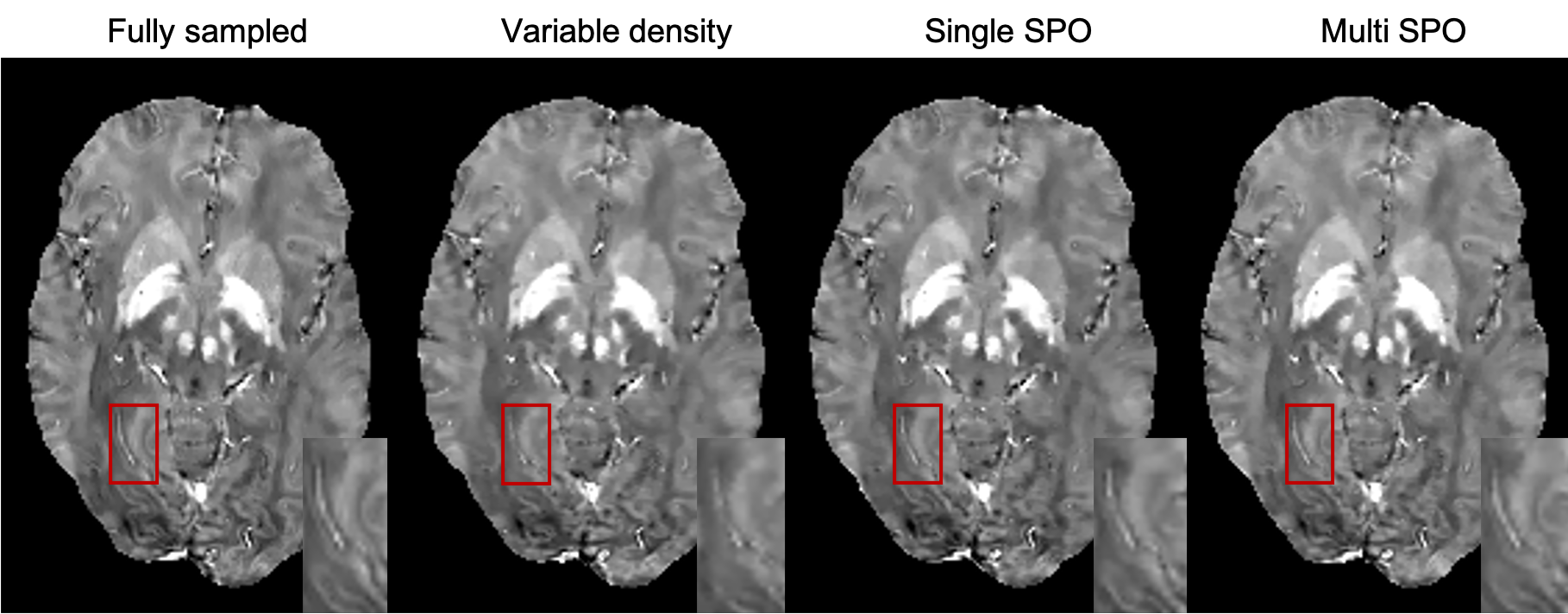}}
  {\caption{QSMs estimated from "deep ADMM + TFF" reconstructions using different sampling patterns. The same veil structures from Fig. 2 appeared bright in QSMs and were progressively improved from manually designed variable density to multi-echo SPO patterns.}}
  \label{fig4}
\end{figure}

\section{Conclusion}
We propose a unified method to optimize MEGRE signal acquisition and image reconstruction. The proposed reconstruction network inserted a recurrent TFF block into a deep ADMM to capture image evolution dynamics along echo time. The proposed SPO block extended the single-echo LOUPE-ST to multi-echo regime. Experimental results showed superior performance for both reconstruction and multi-echo sampling pattern compared to other methods and patterns.

\newpage
%
%
%
\bibliographystyle{splncs04}
\bibliography{samplepaper}

\end{document}